\documentclass{emulateapj}

\usepackage{subfigure}
\usepackage{url}
\usepackage{multirow}
\usepackage{color}
\usepackage{graphicx}
\usepackage{xspace}
\usepackage{natbib}
\usepackage{amsmath}
\usepackage[figuresright]{rotating}

\newcommand{\ccolor}{\textcolor{black}} 
\newcommand{\water}{$\rm H_2O$} 
\newcommand{\methane}{$\rm CH_4$}
\newcommand{\ammonia}{$\rm NH_3$}
\newcommand{\cotwo}{$\rm CO_2$}
\let\cp\citep
\let\ct\citet

\shorttitle{2D Retrievals of Exoplanet Thermal Emission}
\shortauthors{Feng et al.}

\begin{document}


\title{\uppercase{The Impact of Non-Uniform Thermal Structure on the Interpretation of Exoplanet Emission Spectra }}


\author{\textsc{Y. Katherina Feng\altaffilmark{1,2}, Michael R. Line\altaffilmark{3,4,5,6}, Jonathan J. Fortney\altaffilmark{1}, Kevin B. Stevenson\altaffilmark{7,8}, Jacob Bean\altaffilmark{7}, Laura Kreidberg\altaffilmark{7} }, Vivien Parmentier\altaffilmark{8,9}}

\altaffiltext{1}{Department of Astronomy \& Astrophysics, 1156 High Street, University of California, Santa Cruz, CA 95064, USA.}
\altaffiltext{2}{NSF graduate research fellow}
\altaffiltext{3}{NASA Ames Research Center, Mountain View, CA, United States}
\altaffiltext{4}{Bay Area Environmental Research Institute}
\altaffiltext{5}{Hubble Postdoctoral Fellow}
\altaffiltext{6}{School of Earth \& Space Exploration, Arizona State University}

\altaffiltext{7}{Department of Astronomy \& Astrophysics, University of Chicago}
\altaffiltext{8}{Sagan Postdoctoral Fellow}
\altaffiltext{9}{Lunar \& Planetary Laboratory, University of Arizona}

\received{May 5, 2016}
\accepted{July 5, 2016}

\begin{abstract}
The determination of atmospheric structure and molecular abundances of planetary atmospheres via spectroscopy involves direct comparisons between models and data.   While varying in sophistication, most model-spectra comparisons fundamentally assume ``1D'' model physics. However, knowledge from general circulation models and of solar system planets suggests that planetary atmospheres are inherently ``3D'' in their structure and composition. We explore the potential biases resulting from standard ``1D'' assumptions within a Bayesian atmospheric retrieval framework.  Specifically, we show how the assumption of a single 1-dimensional thermal profile can bias our interpretation of the thermal emission spectrum of a hot Jupiter atmosphere that is composed of two thermal profiles. We retrieve upon spectra of unresolved model planets as observed with a combination of  \textit{HST} WFC3+\textit{Spitzer} IRAC as well as \textit{JWST} under varying differences in the two thermal profiles. For WFC3+IRAC, there is a significantly biased estimate of \methane\ abundance using a 1D model when the contrast is 80\%. For \textit{JWST}, two thermal profiles are required to adequately interpret the data and estimate the abundances when contrast is greater than 40\%.  \ccolor{We also apply this preliminary concept to the recent WFC3+IRAC phase curve data of the hot Jupiter WASP-43b. We see similar behavior as present in our simulated data: while the \water\ abundance determination is robust, \methane\, is artificially well-constrained to incorrect values under the 1D assumption.} Our work demonstrates the need to evaluate model assumptions in order to extract meaningful constraints from atmospheric spectra and motivates exploration of optimal observational setups.

\end{abstract}


\section{Introduction} 

\label{intro}

Even a cursory view of images of solar system planets shows us that these planets have complex atmospheres.  It is readily appreciated that not all latitudes and longitudes look alike.  A view of Jupiter at 5 $\mu$m shows bright bands and spots, where, due to locally optically thin clouds, thermal emission can be seen from deeper, hotter atmospheric layers.  Looking at Mars in visible light, we can often see locations obscured by thin cirrus clouds in the atmosphere, and at other locations we can see down to the surface.  These different locations not only appear different to our eyes; the spectra of light that they reflect and emit also differ.  When it is possible to resolve the disk of the planets under study, quite detailed levels of information can be determined:  for instance, changing cloud properties with latitude, different atmospheric temperature-pressure (TP) profiles with solar zenith angle, and compositional differences in updrafts vs.~downdrafts.

However, if a planet is tens of parsecs distant, there is no path to spatially resolving the visible hemisphere (with current technology).  Observers probe the spectra reflected or emitted by the visible hemisphere, but there is generally little hope of assessing how diverse or uniform the visible hemisphere is.  Typically, when comparing observations to the spectra from either self-consistent radiative-convective forward models \cp[e.g.][]{Burrows07c,Fortney08a,Marley12,Barman11}, or from data-driven retrievals \cp[e.g.][]{Madhu10,Line14}, the spectrum, or set of spectra, are generated and aim to represent hemispheric average conditions.  However, while the calculation of such a spectrum, and its comparison to data, is relatively straightforward, it has been unclear how dependent our inferences are for TP profile structure, cloud optical depth, and chemical abundances from this important initial assumption.

Recent work on matching the spectra of some brown dwarfs and directly imaged planets points to problems with the homogeneous atmosphere assumption, with best-fit radiative-convective forward models coming from spectra generated from linear combinations of ``cloudy'' and ``clear'' atmospheres, or atmospheres with weighted areas of ``thick'' and ``thin'' clouds \cp{Skemer14,Buenzli14}.  The variable nature of brown dwarf thermal emission, now well documented over several years via photometry \cp[e.g.,][]{Enoch03,Artigau09,Radigan14} and spectroscopy \cp{Buenzli14,Buenzli15}, also indicates inhomogeneity in the visible hemisphere, with emission that changes due to rotation and/or atmospheric dynamics \cp{Robinson14,Zhang14,Morley14b,zhou2016}.

In the realm of retrievals, could a search through phase space for a best-fit to a measured spectrum lead to well-constrained yet biased or incorrect constraints on atmospheric properties when we assume planet-wide average conditions?  This seems like a real possibility, and one well worth investigating in a systematic way.  With the advent of higher signal-to-noise spectroscopy from the ground \cp{Konopacky13} and the coming launch of the \textit{James Webb Space Telescope} (\textit{JWST}), which will deliver excellent spectra for many planets over a wide wavelength range, we aim to  test the 1D planet-wide average assumption systematically.  We want to furthermore determine, when the data quality is high enough, if we can justify a more complex inhomogeneous model.

Recently, \ct{Line15} investigated for transmission spectra how the signal of high atmospheric metallicity inferred under planet-wide average conditions can be mimicked by a uniform lower metallicity together with a high cloud over a part of the planet's terminator.  Our work here is on thermal emission and is entirely complementary.  We take the first step in characterizing how a diverse visible hemisphere may impact atmospheric retrievals.  

Our paper is organized as follows: Section \ref{method} describes the setup, retrieval approach, and methodology.  In Section \ref{results}, we describe our findings. In Section \ref{w43b}, we present the application to WASP-43b. We discuss our results in Section \ref{discussion} and conclude with future expansions.

\section{Methodology}\label{method}
\subsection{Setup}
\begin{figure}[ht!]
\includegraphics[width=0.45\textwidth]{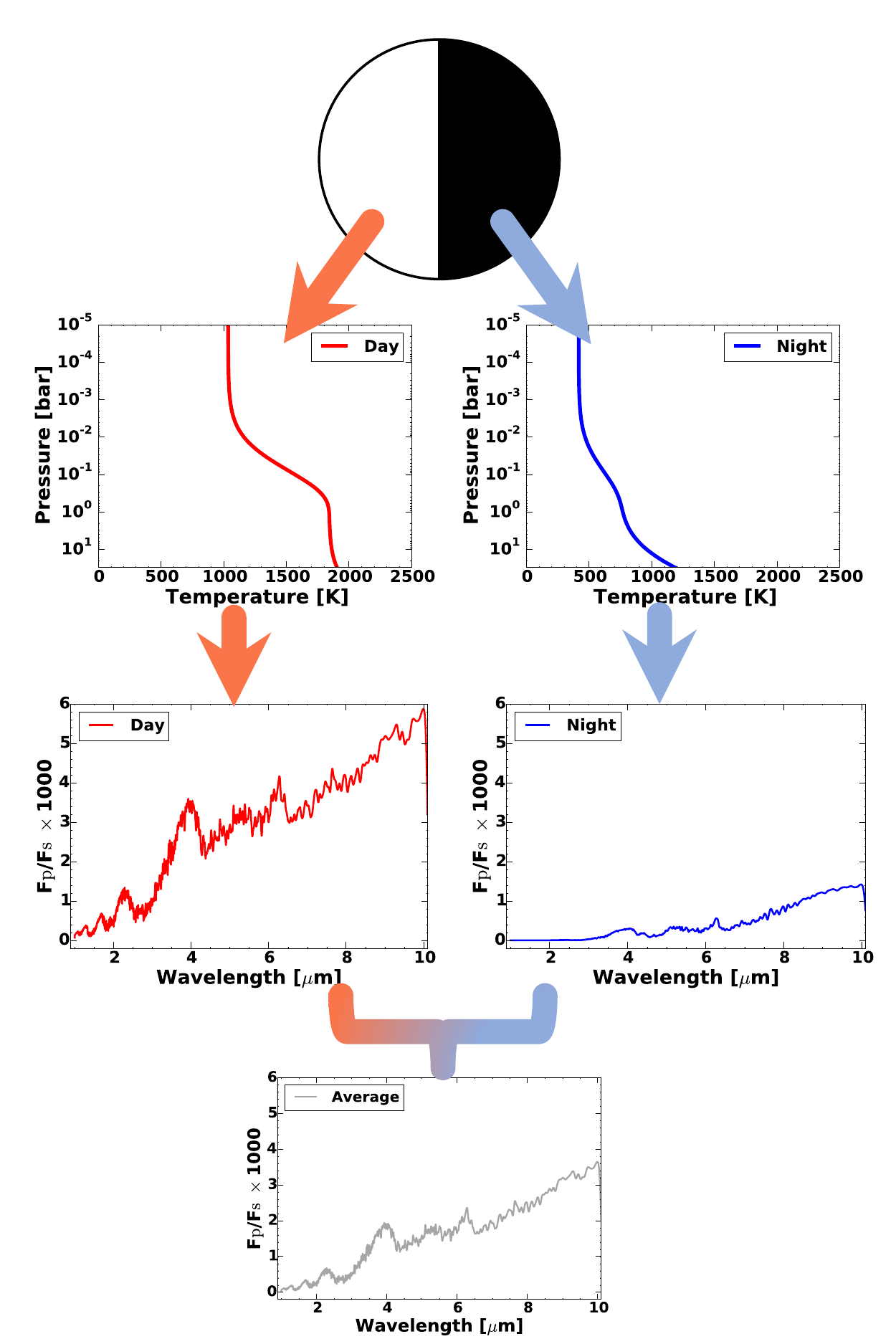}
\caption{\label{fig:schematic}
Schematic demonstration of our set up. We assume a planet with two equally weighted thermal structures with a cloud-free atmosphere of uniform composition. The fluxes from both thermal profiles are then averaged to create the disk integrated spectrum \ccolor{upon which we perform the retrievals}.}
\label{demo}
\end{figure}

We present a simple case to illustrate the impact of a planet's spatially varying thermal structure on retrievals.   The model setup features two different TP profiles, equally weighted in surface area,  in a cloud-free atmosphere with planet-wide uniform abundances. This case is relevant to two simple kinds of atmospheres. One is a ``checkerboard'' atmosphere of equal-area hotter and colder areas, with applicability to brown dwarfs and imaged planets.  Another is a transiting planet with a hot day side and cold night side, as viewed at one-quarter or three-quarter phase, meaning half-day and half-night. The equal-weighting average allows for symmetry in viewing geometry and limb-darkening effects. Each TP profile generates emitted fluxes at the top of the atmosphere. The observed spectra result from the average of the fluxes. From these averaged spectra, we generate data as observed with typical \textit{Hubble Space Telescope}  (\textit{HST}) Wide Field Camera 3 (WFC3)+\textit{Spitzer} Infrared Array Camera (IRAC) and  \textit{JWST} modes.  We then perform atmospheric retrievals on these synthetic data assuming either a single profile (1TP) or two profiles (2TP). Figure \ref{fig:schematic} shows the setup. 

As an initial investigation, we primarily explore the role that temperature contrast has in biasing the retrieval results, specifically on hot Jupiters. The TP profiles are offset at the top of the atmosphere from each other by a factor (i.e. contrast) defined as 
\begin{equation}
1-\frac{T_{\rm TOA,c}}{T_{\rm TOA, h}}
\label{toa}
\end{equation}
where $T_{\rm TOA,c}$ and $T_{\rm TOA,h}$ are the top-of-atmosphere (TOA) temperatures for the cold (``night") and hot (``day") TP profiles, respectively.  Under different observational setups, we (1) determine the biases in the atmospheric abundances when one global TP profile is assumed for a planet that is actually composed of two TP profiles and (2) quantitatively determine the justification for the inclusion of a second TP profile within a nested model hypothesis testing framework \citep[e.g.,][]{trotta08,cornish07}. In what follows, we describe the necessary tools to accomplish these tasks.    

\subsection{Modeling Tools}
The thermal infrared radiative transfer model we use is described in detail in \ct{Line2013}. It numerically solves the thermal infrared radiation problem for a plane-parallel atmosphere with absorption, emission, and no scattering given a TP profile and uniform-with-altitude gas abundances.  We consider absorption due to  \methane, \cotwo, CO, \water, \ammonia, He, and  H$_2$. The molecular abundances for generating the synthetic spectra are chosen to be in rough agreement with solar elemental abundances in thermochemical equilibrium at a representative photospheric pressure ($\sim$100 mbars) along the prescribed thermal profile.  When computing the spectra for two TP profiles, we assume the dayside abundances for both, consistent with expectations from horizontal mixing \citep[e.g.,][]{CS06,Agundez2014}.   The opacity database is described in \citet{Freedman2014}. 

We set and retrieve for the temperature profiles using the \citet{parmentier14} 5-parameter prescription (two visible opacity parameters ($\log \gamma_1$, $\log \gamma_2$), partitioning between the two visible streams ($\alpha$), infrared opacity ($\log \kappa_{IR}$), and the fraction of absorbed incident flux ($\beta$); see Equations 13, 14 in \ct{Line2013} and Table \ref{params}). The internal temperature, T$_{\rm int}$, is an additional parameter we specify, but it is not one of the retrieved quantities. We fix T$_{\rm int}$ to 200K \citep{Guillot10} which prevents the TP profiles from ever reaching 0 K.  Given the molecular abundances and thermal structures, we use four point Gaussian quadrature to compute the full disk-integrated spectrum for the day and night profiles separately. By taking the average of the ``hemispheres" or ``checkerboard patches"  and dividing by a stellar spectrum, we generate the planet-to-star flux ratios. Taking the average of the disk-integrated fluxes is equivalent to weighting each profile by the same area, thus invoking the same limb-darkening effects. Note that this need not be true in the case of hot-spot or ``crescent phase" models, where there is asymmetry in limb darkening, which we will investigate in a later publication.

The high-resolution model spectra are then appropriately convolved and interpolated to the ``observational" wavelength grid.  Poisson noise (no systematic noise is included) is then added to each data point.  For the \textit{HST} WFC3+\textit{Spitzer} IRAC setup, we assume error bars representative of current observations (e.g., 35 ppm error bars at 0.035 $\mu$m resolution, R$\sim$40, for WFC3, and 70 ppm error bars for the \textit{Spitzer} IRAC 3.6 and 4.5 $\mu$m channels).  For the \textit{JWST} observational setup, we use the noise model described in \citet{greene2016}, covering 1-11 $\mu$m and combining modes from the NIRISS, NIRCam, and MIRI instruments.  This noise model simulates the uncertainties obtainable with the observation of a single transit or secondary eclipse for a hot Jupiter. We adopt planet and stellar parameters for the HD 189733 system (Table \ref{tab:Table1}).  

We use an atmospheric retrieval approach to explore the biases introduced in assuming a single TP profile for a spectrum composed of two separate TP profiles and the detectability of multiple profiles.  Much of the thermal infrared retrieval machinery is based on the CHIMERA retrieval suite already described in \citet{Line2013} and \citet{Line14} and subsequently applied in \citet{Kreidberg14} and \citet{stevenson14}. However, for the Bayesian inference problem, rather than using the differential evolution Monte Carlo approach, we use the {\tt multinest} algorithm \citep{MN} as implemented with the {\tt pymultinest} routine \citep{PMN} because it not only has the ability to produce posterior samples, but it also efficiently computes the Bayesian evidence, or the integral of the posterior over the parameter volume.   The Bayesian evidence is required for model comparison and selection, and it is a numerical encapsulation of the balance between goodness-of-fit and a model's simplicity. It can be thought of as the more rigorous computation of the commonly used Bayesian Information Criterion (BIC). For application of the {\tt multinest} algorithm and model selection to exoplanet spectra, we refer the reader to \citet{benneke13}, \citet{waldmann15a}, \citet{waldmann15b}, and \citet{Line15}. For a summary of Bayesian model selection and evidence computation, we refer the reader to \citet{trotta08} and \citet{cornish07}.   

From the synthetic model spectra, we aim to determine the constraints on the uniform-with-altitude abundances for \methane, \cotwo, CO, \water, \ammonia, and both of the TP profiles.  We assume the same 4 ``shape" parameters (two visible opacity parameters, partitioning between the two visible streams, and the infrared opacity) for both TP profiles but allow for a different ratio of the absorbed-to-incident flux (e.g., some combination of albedo and redistribution), represented by parameter $\beta$ in Table \ref{params}. \ccolor{$\beta$ also acts as a multiplicative factor between the two contrasting profiles, as illustrated by our definition of contrast in Equation \ref{toa}}. This leads to a total of 11 free parameters for the 2TP model and 10 for the 1TP model. We assume uniform-in-log$_{10}$ priors for the 5 gas volume mixing ratios ranging from -12 to 0 and top-of-atmosphere temperatures ranging from zero to twice the irradiation temperature.

\def\arraystretch{1.5}
\begin{deluxetable}{lr|lr}[H]
\tablecaption{Model Parameter values \label{params}}
\tablewidth{0.35\textwidth}
\tabletypesize{\scriptsize}
\tablehead{Parameter & Value & TP Parameter & Value}
\startdata
\label{tab:Table1}
$R_p$ ($R_{\rm J}$)&  1.138 & $\log \gamma_1$ & -1\\
$R_*$ ($R_{\odot}$)&0.756 & $\log \gamma_2$ & -1\\
$T_*$ (K) & 5040 & $\log \kappa_{\rm IR}$ & -1\\
$a$ (AU)  &0.031 & $\alpha$ & 0.5\\
$T_{\rm int}$ (K) &200 & $\beta_{\rm day}$ & 1\\  
$\log(g)$ (cm s$^{-2}$)  &3.34 & $\beta_{\rm night}$ & 0.2\\  %
$\log f_{\rm H_2O}$ & -3.37 & &0.4\\
$\log f_{\rm CH_4}$ & -9 & &0.6\\
$\log f_{\rm CO}$ &-3.7& & 0.8\\
$\log f_{\rm CO_2}$ & -9& &\\
$\log f_{\rm NH_3}$  &-9& &\\ 
\enddata
\tablecomments{Nominal system and TP shape parameters used to generate our synthetic spectra. Stellar and planetary parameters are based on the HD 189733 system. For definitions of the TP parameters, see \citet{Line2013}. Solar proportion Hydrogen and Helium are assumed to make up the remaining gas abundance.}
\end{deluxetable}

\begin{figure*}[ht!]
\begin{center}
\includegraphics[width=\textwidth]{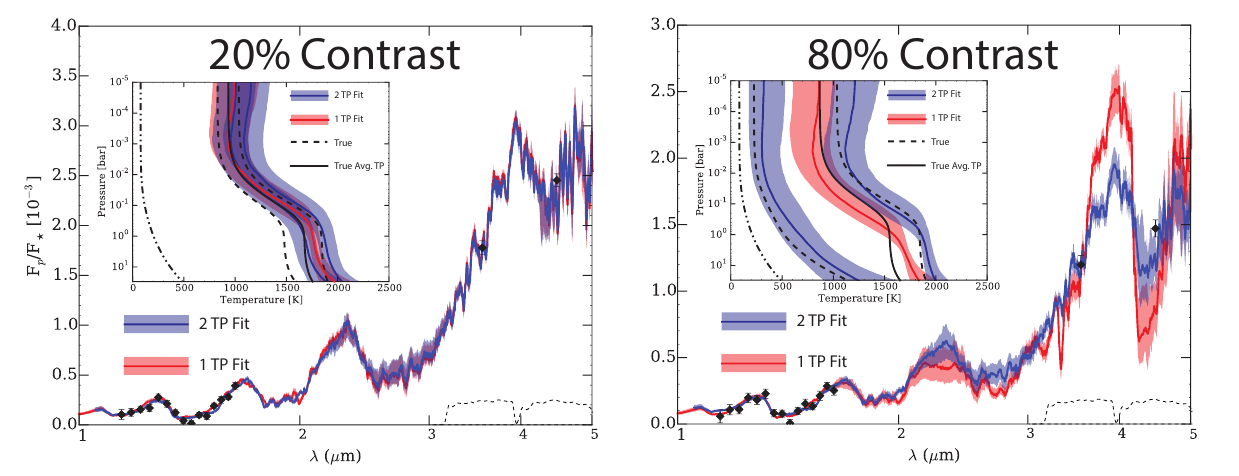}
\end{center}
\caption{\textit{HST} WFC3 + \textit{Spitzer} IRAC 1TP vs. 2TP fit and temperature profiles (insets) retrieval summary.  The left panel shows the results for the low (20\%) contrast while the right shows the results for high (80\%) contrast.  The data simulated with 2 TP profiles are shown as the black diamonds with error bars (WFC3 between 1 and 2 $\mu$m and the \textit{Spitzer} IRAC points at 3.6 and 4.5 $\mu$m).  The fits and temperature profiles are summarized with a median (solid line) and 68\% confidence interval (spread) generated from 1000 randomly drawn parameter vectors from the posterior.    Red corresponds to the fits/temperature profiles resulting from a single profile fit, while blue represents the result of including two temperature profiles in the retrieval. The black dashed lines in the temperature profile insets are the two TP profiles used to generate the simulated data (i.e., the ``true" TP profiles).  For comparison, we also include the flux-averaged TP profile ($T_{\rm avg}^4 = \frac{1}{2}(T_{\rm day}^4 + T_{\rm night}^4)$), shown as the solid black line in the insets. The dot-dashed TP profile is the coldest profile permitted by the model: a non-irradiated cooling profile governed by the 200K internal temperature. \ccolor{By eye, the 1TP vs. 2TP performances at 20\% contrast are comparable. Based on the Bayesian evidence, the detection of the second profile is not significant ($<\ 0.1 \sigma$). At 80\% contrast, the two retrieved spectra are visibly different.  The second profile is detected to $2.4\sigma$ significance.} }
\label{wfc3_spec}
\end{figure*}

\begin{figure*}[ht!]
\begin{center}
\includegraphics[width=\textwidth]{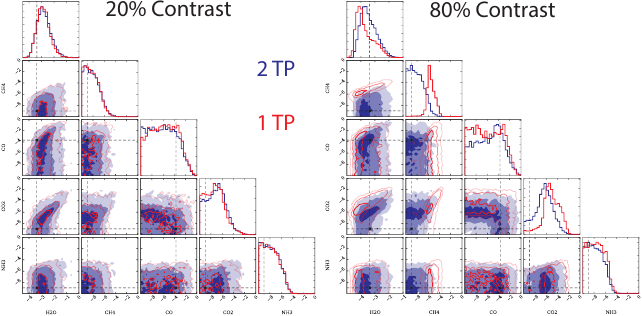}
\end{center}
\caption{Summary of the posterior probability distributions of the molecular abundances for the low (20\%, left) and high (80\%, right) contrast cases under the \textit{HST} WFC3+\textit{Spitzer} IRAC observational scenario. The red and blue 1- and 2-D histograms correspond to  1TP and 2TP scenarios. The dashed lines in the 1-D histograms and intersection of the dashed lines in the 2-D histograms are the true molecular abundances used to generate the synthetic data. \ccolor{The detection significance of the second profile from the 2TP retrieval is $<\ 0.1 \sigma$ at 20\% contrast, and the posterior distributions show that invoking a second profile did not improve our abundance estimation. At 80\% contrast, where the detection significance is $2.4\sigma$, we still note the similarities in the posterior distributions for most species.  However, in the case of \methane, the 1TP approach, bound by the radiative transfer properties of one profile, overestimates both its abundance and the precision. When we include a second profile, we are able to recover a more realistic and representative distribution for the \methane\, abundance. }}
\label{wfc3_08}
\end{figure*}

\begin{figure*}[ht!]
\begin{center}
\includegraphics[width=\textwidth]{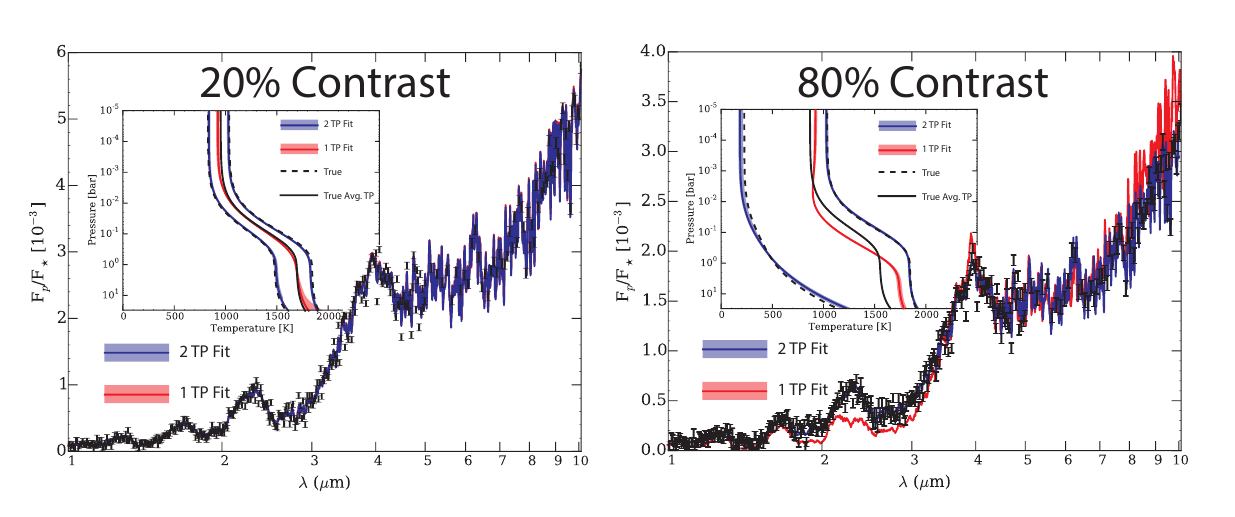}
\end{center}
\caption{\textit{JWST} 1TP vs. 2TP fit and temperature profiles (insets) retrieval summary.  The left shows the results for the low (20\%) contrast while the right shows the results for high (80\%) contrast.  The data simulated with 2 TP profiles are shown as the black error bars.  The fits and temperature profiles are summarized with a median (solid line) and 68\% confidence interval (spread) generated from 1000 randomly drawn parameter vectors from the posterior.    Red corresponds to the fits/temperature profiles resulting from a single TP profile fit, while blue represents the result of including two temperature profiles in the retrieval. The black dashed lines in the temperature profile insets are the two TP profiles used to generate the simulated data (e.g., the ``true" TP profiles).  For comparison, we also include the flux-averaged TP profile ($T_{\rm avg}^4 = \frac{1}{2}(T_{\rm day}^4 + T_{\rm night}^4)$), shown as the solid black line in the insets. \ccolor{At 20\% contrast, while the retrieved fits appear similar, we find that the second TP profile is detected to $\sim 5\sigma$ significance. At 80\% contrast, the 1TP retrieved spectra poorly fit the data, especially at $2-3\ \mu$m and at longer wavelengths. }}
\label{jwst_spec}
\end{figure*}

\begin{figure*}[ht!]
\begin{center}
\includegraphics[width=\textwidth]{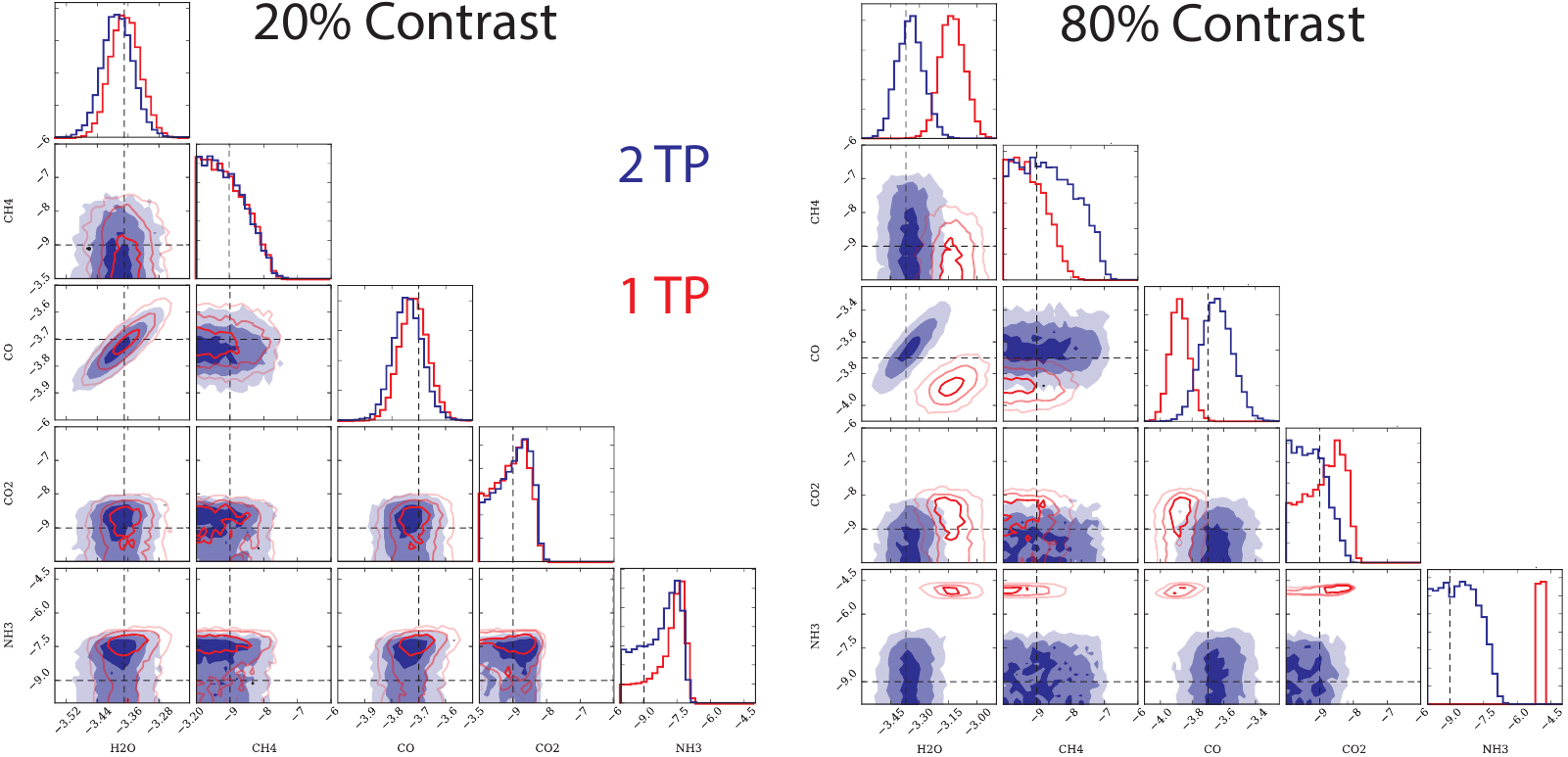}
\end{center}
\caption{Summary of the posterior probability distributions of the molecular abundances for the low (20\%, left) and high (80\%, right) contrast cases under the \textit{JWST} observational scenario. The red and blue 1- and 2-D histograms correspond to  1TP and 2TP scenarios. The dashed lines in the 1-D histograms and intersection of the dashed lines in the 2-D histograms are the true molecular abundances used to generate the synthetic data. \ccolor{When the contrast is 20\%, the second profile is detected to $\sim 5\sigma$. When the contrast is 80\%, the second profile is detected to $>\ 20 \sigma$. We see that, at higher contrasts, the 1TP retrieval case is a poor representation of the abundances. We also note the over-constraint of \ammonia\, under the 1TP prescription. This behavior is analogous to the \methane\, abundance inference using one profile that we saw with WFC3+IRAC data. Once a second profile is included, we recover the true abundance of \ammonia.}    }\label{jwst_08}  
\end{figure*}

\begin{deluxetable}{c|cc|cc}[H]
\tablecaption{Retrieval Results and Bayesian Model Evidence for 2nd TP profile \label{sigmatable}}
\tablewidth{0pt}
\tabletypesize{\scriptsize}
\tablehead{& \multicolumn{2}{c|}{WFC3} 	&  \multicolumn{2}{c}{\textit{JWST}}		\\
Contrast& ln($B$)\tablenotemark{a} &$\sigma$\tablenotemark{b}  & ln($B$) & $\sigma$ \\
term& & 2nd TP & &2nd TP
}
\startdata
0.2	& -1.06$\pm$0.68	&		$<$0.1	&	12.93	&	5.44\tablenotemark{c}	\\
0.4	&	1.12$\pm$0.56	&2.05$\pm$0.37	&	274.8	&	$>$20	\\
0.6	&2.49$\pm$2.10	&		2.54$\pm$1.00	&	967.9	&	$>$20	\\
0.8	&	1.77$\pm$0.75	&	2.41$\pm$0.34	&	1836	&	$>$20	\\
\hline
1TP	&	-1.26	&	$<0.1 $ &	-2.92	&	$<0.1$

\enddata
\tablecomments{The last row, ``1TP", reports the case for which we generated the spectrum with one TP profile and retrieved for two. For both observational setups, in this scenario, a 2nd TP profile is not favored. Contrast term is $1-\beta_{\rm night}$ (see Table \ref{params}), and ``$\sigma$ 2nd TP'' is the detection significance of the 2nd TP profile.  } 
\tablenotetext{a}{Bayes factor, calculated as the difference in the natural log of the evidence between the larger model (2TP) and the smaller model (1TP).}
\tablenotetext{b}{We consider a $>3.6\sigma$ detection to be strong  \citep[][Table 2]{trotta08} }
\tablenotetext{c}{Using a different noise instance, we find a 4.2$\sigma$ detection of the 2nd profile. While ln($B$) changed, the 2nd TP is still detected robustly. \ccolor{We also calculate the BIC for the noise instance with $5.4\sigma$ detection significance. Our $\Delta$ BIC = 23, which is above the threshold ($\Delta$ BIC $>$ 10) for strong evidence against the model with the larger BIC (in our case, the 1TP scenario). It also corresponds to $5\sigma$ detection significance, consistent with the Bayesian evidence result. Small differences in $\chi^2$ are magnified if there are many points, as with \textit{JWST} data.}}
\end{deluxetable}

\section{Results}\label{results}
We present our retrieval results on the synthetic spectra simulated with \textit{HST} WFC3+\textit{Spitzer} IRAC and \textit{JWST}. For each telescope combination, we produce spectra for four levels of contrast between the two TP profiles (see Equation \ref{toa}) of 0.2, 0.4, 0.6, and 0.8. For each spectrum, we perform a 1TP and a 2TP retrieval. We also test the inclusion of two TP profiles in the retrieval when only one profile was used to generate the spectrum and synthetic data. We compare the performance of the two models by the Bayes factor, summarized in Table \ref{sigmatable}. Based on the retrievals, we can explore the biases resulting from retrieving for a single TP profile when the spectrum is generated with two. We also quantify the detectability of a second TP profile as contrast changes.  

We summarize the retrieval results comparing the 1TP and the 2TP retrievals for only the extreme contrasts, 0.2 and 0.8. In our figures, we also include a flux weighted (averaging $T^4$) profile for each contrast to guide the eye when interpreting the 1TP retrievals.  One would expect a single representative TP profile to closely match the flux weighted profile.  The gas abundance retrievals are summarized with a pairs-plot showing both the 1D and 2D marginalized posteriors (Figure \ref{wfc3_08} for \textit{HST}+\textit{Spitzer}, and Figure \ref{jwst_08} for JWST). The TP profiles and spectra are summarized with a median and 1-sigma spread reconstructed from 1000 randomly drawn posterior samples (Figure \ref{wfc3_spec} for \textit{HST}+\textit{Spitzer}, Figure \ref{jwst_spec} for JWST).

We note that, because we knew {\it a priori} (from test simulations) that the detection significance would be marginal for a 2nd TP profile within the \textit{HST}+\textit{Spitzer} setup, we tested their robustness by performing the retrievals and nested model comparison on six noise instances per contrast setup.  At low detection values (less than $\sim 3\sigma$), the exact detection significance is very sensitive to a particular random noise instance. Thus, in Table \ref{sigmatable}, we show a mean value and error on the results for the \textit{HST}+\textit{Spitzer} observational setup.  This is not an issue for the \textit{JWST} observational setup as the detection significances are always above a significant threshold.

\subsection{Findings for Simulated \textit{HST} WFC3+\textit{Spitzer} IRAC Observations}
Figure \ref{wfc3_spec} and \ref{wfc3_08} summarize the results for the 0.2 and 0.8 contrast cases.  The left panel of Figure \ref{wfc3_spec} shows the retrieved TP profiles and model spectra for the low contrast (0.2) scenario.  The spectra are nearly indistinguishable. This results in our inability to robustly distinguish a 2nd TP profile as the 68\% confidence envelopes for each of the two TP profiles strongly overlap with each other and with the error envelope for the single TP profile. The retrieved molecular abundance posteriors (Figure \ref{wfc3_08}, left panel) are also nearly indistinguishable between the 1 and 2TP cases.  Unsurprisingly, the nested model comparison results in a non-detection for a 2nd TP profile  in the 0.2 contrast scenario (Table \ref{sigmatable}). In fact the Bayes factor, $B$, is less than 1 ($\ln B <0$) suggesting that the improvement in the spectral fit is outweighed by the increased model complexity.   

At higher contrast (0.8), there is a greater deviation in the retrieved model spectra at longer wavelengths ($>2 \mu$m). The 1TP spectra have to contort themselves with a strong peak-to-trough ``N''-shaped feature between 3 and 5 $\mu$m in order to fit the two IRAC points. The broadband integration over wavelength does not allow us to tell the difference between the two scenarios.  The day and night TP profiles, in contrast to the 0.2 scenario, are widely separated outside of their 68\% confidence intervals.  However, the 68\% confidence envelope of the 1TP profile largely encompasses the flux-weighted TP profile, especially over the range where the observations probe \citep[between 1 and 0.01 bars, ][]{stevenson14}. We note that the fixed internal temperature of 200 K sets a lower limit for the night side profile, while the retrieved profiles serve as an upper limit. We saw this by examining the histogram of retrieved temperatures at a certain pressure ($4$ mbar); the distribution is unbounded but consistent with the coldest permitted temperature. While the detection significance for the 2nd TP profile in Table \ref{sigmatable} is higher than in the 0.2 contrast case, it is still not considered significant. However, it makes sense that an increase in contrast should result in higher evidence for a 2nd TP profile. 

Perhaps the most striking find in this high-contrast scenario is the strong differences in the molecular abundance constraints, in particular that of \methane. While the 2TP scenario (the ``true" model) results in an upper limit on the methane abundance, as expected given the low (non-detectable) input value used,  the 1TP profile scenario results in a strong methane constraint.  This strong constraint, however, is several standard deviations away from the true input abundance. In essence, assuming only one TP profile results in an artificial constraint on the methane abundance. This is a key result that we would like to highlight.  The narrow constraint is due to the high sensitivity of the fit (due to the topology of the hyper-dimensional likelihood volume) to small changes in the \methane\, abundance within the 1TP setup, very much like what would happen if one were to fit a constant to linear data. This is largely driven by the IRAC 3.6 $\mu$m point. \ccolor{In terms of the other species, we find that the distribution for \cotwo\, is sensitive to the noise instance of the data points (especially $4.5\ \mu$m), and performs more closely to the true value under a 2TP retrieval depending on the noise instance.} 

\ccolor{The dramatic change in emission from $3-5\mu$m in the 1TP-retrieved spectra for WFC3+IRAC (Figure 2b), showing strong emission and then absorption, merited additional modeling to investigate its cause. To investigate these prominent absorption features present across the IRAC bands we performed a 1TP retrieval where all the abundances were fixed to their true values, to better understand the role of TP profile shape in generating the spectrum.  With this reduced parameter space, it was more readily apparent that the retrieved TP profiles featured a significant temperature gradient -- 1000 K -- that spanned $\sim 700-1700$ K over a relatively narrow pressure range ($0.01 - 1$ bar). These large differences in temperatures probed naturally leads to the striking features (strong emission and absorption) seen in the spectrum, while a more isothermal profile would yield more muted contrasts in emission.}

Finally, from Table \ref{sigmatable}, we find that the detection of a 2nd TP profile is below what is commonly considered significant (3-4$\sigma$), especially when considering the uncertainties, but that all contrasts greater than 0.2 are more justified in including the 2nd TP profile.  The marginal detections are a result of the complex interplay between the intrinsic temperature contrast, wavelength coverage, and feature signals-to-noise. Furthermore, as a sanity check, we find evidence {\it against} ($\ln B < 0$) the inclusion of a 2nd TP profile when only one is used to create the spectra.

\subsection{Findings for Simulated JWST data}
Figure \ref{jwst_spec} and \ref{jwst_08} summarize the findings for 0.2 and 0.8 contrast cases. We find that for low contrast (0.2) there is not a significant bias in the retrieved molecular abundance when using one TP profile, and that the retrieved single TP profile matches the flux-averaged TP profile quite well. However, we still find a significant detection ($>5\sigma$) of a 2nd TP profile. This suggests that fit with the single TP profile is not quite as good as the fit with two, though apparently indistinguishable by eye, even when taking into account the Occam's penalty \citep[Table \ref{sigmatable}; ][page 49]{gregory05}.  

The situation changes, however, for large contrasts (0.8).  The 1TP fit is noticeably worse between $\sim 1.6 $ and $3.3 \mu$m, and then again at the longest wavelengths (Figure \ref{jwst_spec}). The shape of the spectrum is different enough with two TP profiles that a model with single TP profile simply cannot accommodate. Because WFC3+IRAC do not cover $2-3\ \mu$m, we would not have known that this range is sensitive to the large TP contrasts in our particular toy atmosphere. Thanks to \textit{JWST}’s wavelength coverage, we see that, at large contrasts, a second profile is needed, and this profile is detected to $>\ 20 \sigma$. Furthermore, the 1TP model results in significant abundance biases.  The \water\, abundance is much higher (relative to the uncertainty) than the truth, CO is slightly underestimated, and ammonia off by $\sim$ 4 orders of magnitude, with an artificially small uncertainty. This is a cautionary note that small uncertainties on parameter values should be taken with a grain of salt if a model is inadequately fitting the data. \ccolor{This behavior is analogous to the \methane\, abundance inference using one profile that we saw with WFC3+IRAC data. Once a second profile is included, we recover a distribution representative of the true abundance of \ammonia.}

For the remaining contrast cases (0.4 and 0.6), we find overwhelming evidence ($>20\sigma$) for the presence of a 2nd TP profile. We also find, as expected, that there is little evidence for a 2nd TP profile from an object with only one TP profile.  All of this taken together suggests that \textit{JWST} observations of thermal emission spectra will be {\it extremely} sensitive to the presence of multiple TP profiles (given reasonable observational assumptions).


\section{Application to WASP-43 b} \label{w43b}
\begin{figure*}[ht!]
\includegraphics[width=\textwidth]{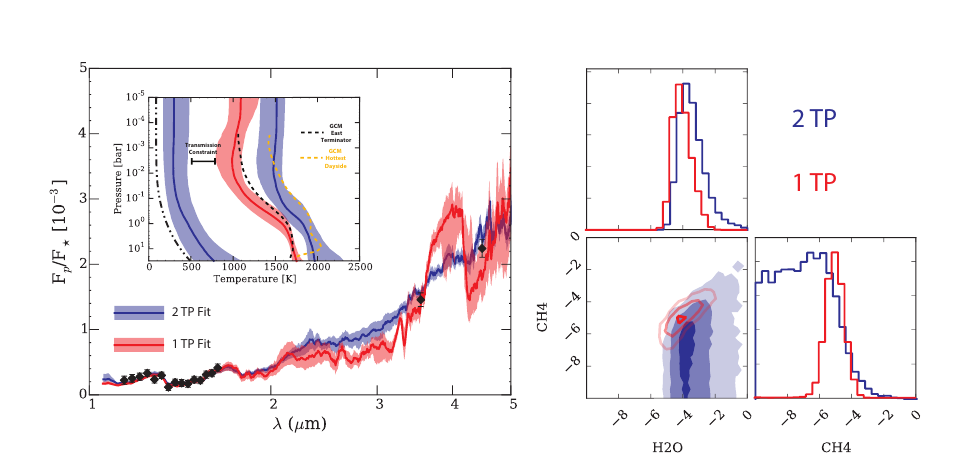}
\caption{Summary of the 1 TP vs. 2 TP retrievals on the \textit{HST} WFC3 + \textit{Spitzer} IRAC observations of WASP-43b. In the left panel, the data are shown as the black diamonds with error bars (WFC3 between 1 and 2 $\mu$m and the \textit{Spitzer} IRAC points at 3.6 and 4.5 $\mu$m).  The fits and temperature profiles (inset) are summarized with a median (solid line) and 68\% confidence interval (spread) generated from 1000 randomly drawn parameter vectors from the posterior.    Red corresponds to the fits/temperature profiles resulting from a single TP profile fit, while blue are a result of including two temperature profiles in the retrieval. The dot-dashed TP profile is the coldest profile permitted by the model: a non-irradiated cooling profile governed by the 200K internal temperature. At two sigma, the retrieved night-side TP profile is consistent with the coldest permitted profile, suggesting that the retrieved night-side temperatures are an upper limit. We also show GCM derived TP profiles for the east terminator (black dashed) and dayside (purple dashed). The single TP profile fit matches the east terminator GCM profile well, while the dayside TP in the 2TP fit matches the GCM derived dayside TP profile reasonably well. The ``scale height" temperature retrieved from the WASP-43b transmission spectra \citep{Kreidberg14} is shown as the horizontal error bar. This temperature assumes an isothermal profile seen in transmission. Finally, the water and methane abundance posteriors are shown in the right panel. For simplicity, we do not show the posteriors of the other molecules whose abundances were retrieved (\ammonia, CO, \cotwo). Note the water abundance here seems invariant under the 1-(red) or 2-(blue) TP assumptions, but the methane abundance is artificially well-constrained when assuming only 1 TP. Approximate thermochemical equilibrium molecular abundances at 1700K, 0.4 bars (dayside photospheric conditions) with solar elemental composition are shown with the dashed lines.      }
\label{w43fig}
\end{figure*}

As an application to real observations, we test our two TP profile assumption on the well-characterized hot Jupiter WASP-43b. WASP-43b was observed as part of a large \textit{HST} Legacy program (PI Jacob Bean) with WFC3 providing 3 primary transits, 2 secondary eclipses, and 3 full spectroscopic phase curves \citep{Kreidberg14,stevenson14}. Such phase curve observations provide a glimpse into the 3-dimensional structure of a planet as different wavelengths probe different atmospheric pressures and the different phases probe different planetary longitudes. These published results were interpreted (using CHIMERA, the same model used here) assuming a single TP profile representation for each spectrum at every phase. We now know from our synthetic tests above that, for objects with strong day-night contrasts (as WASP-43b possesses), assuming a single TP profile for a single disk-integrated spectrum may result in biased abundances.   Motivated by recent full phase IRAC 3.6 and 4.5 $\mu$m observations of WASP-43b (Stevenson et al., in prep), we decided to revisit interpretation of the spectral energy distribution of non-secondary or primary eclipse phases within our newly developed two-TP profile framework. For an initial exploration, we focus on the first quarter \textit{HST} WFC3+\textit{Spitzer} IRAC data (eastern hemisphere).  This phase represents exactly the geometry explored in above examples:  half ``day",  half ``night". We utilize the same forward model and retrieve for the identical set of molecules as on our simulated data. Figure \ref{w43fig} summarizes the relevant results. In addition to the first quarter, we examined the third quarter (western hemisphere) as well as the day side emission data.

For the first quarter, as in our synthetic WFC3+IRAC example, we find evidence for a bias in the \methane\ abundance.  Assuming a single TP profile forces a solution that results in an overly well constrained methane abundance, an abundance that is a few sigma larger than anticipated from solar composition gas in thermochemical equilibrium at dayside photospheric conditions (1700K, 400 mbars). Once again, one would not expect such a good constraint given that these particular observations only provide a single measurement, the 3.6 $\mu$m band, on a methane absorption feature.  However, as in the above synthetic examples, we find that the water abundance is robust against the 1 vs. 2 TP profile assumption.  This is because water is primarily constrained at shorter wavelengths where the impact of including a second TP profile is minimized.

We determine the justification for the inclusion of a second TP profile by comparing the Bayesian evidence for a model with and without the second profile. Upon doing so, we determine that the second TP profile is justified at the 3.3 $\sigma$ level (just below what would be considered ``strong" on a Jeffery's scale \citep{trotta08}). The Bayes factor is ln$(B) = 3.99$. While this is not the strongest of detections, when combined with the \methane\ bias, it warrants the inclusion of the second TP profile.  We also find that the two retrieved profiles match remarkably well with the hemispheric TP profiles retrieved for the day side and night side spectra presented in \citet{stevenson14} as well as the hottest day side profile from the General Circulation Model (GCM) in \citet{kataria2015}. The projected spectra between 3.8 and 5 $\mu$m show the strongest divergence between the 1- and 2-TP profile fits, followed by wavelengths between 2.2 and 3.5 $\mu$m   .  Future higher resolution observations should focus on these spectral regions to boost the detection level of a ``2nd TP profile".

When we investigated the day side emission data, the one and two TP profile scenarios yielded similar results, consistent with what we saw when the contrast is low between two profiles: on the day side, a second, cooler profile is not needed to explain the data.

We then examined the role of multiple TP profiles for the the third quarter. We found that the second TP profile is not justified by the data (2.7$\sigma$). Like the first quarter single TP profile fit, we find a well-constrained methane abundance using one profile. However, after including a second profile, the methane posterior remained constrained unlike in the case of the first quarter.

While the first and third quarters seem in conflict with regards to the impact of a second TP profile, the full phase curve for WASP-43b shows asymmetry, suggesting that the third quarter is not the exact ``opposite'' to the first quarter in the sense of contrast.   When it comes to seeking trends or consistency throughout the phase, we have to be wary and investigate more thoroughly to differentiate what is truly representative of the atmosphere and what is the artificial manifestation of e.g., the sensitivity to the slope between band-integrated data points.

The full phase curve data of WASP-43b continue to serve as a benchmark data set in the context of the 3D nature of planets and push us to better our model interpretations, which are especially important for future exoplanet characterization observatories.

\section{Discussion and Summary} \label{discussion}

The interpretation of exoplanet spectra is complex; the conclusions we draw about the composition, thermal structures, and other properties of exoplanet atmospheres strongly depend on our model assumptions. In this pilot study, we explored the biases in thermal structure and molecular abundances as a result of the commonly used assumption of ``1D" on the interpretation of transiting exoplanet emission spectra.  We generated spectra from a simple ``2D" setup of a planetary hemisphere composed of two thermal profiles,  representative of either a ``checkerboard" hemisphere, which may physically correspond to a planet peppered with various convective cells, or a ``half-and-half" planet, similar to simultaneously observing a hot day side and cooler night side. We then applied commonly used atmospheric retrieval tools under the assumption of a single 1D homogeneous hemisphere to one that is inherently ``2D". 

Within this setup, we explored how the biases in the abundances and 1D thermal profile are influenced by varying degrees of ``contrast" between the two TP profiles for two different observational situations. We found that, for current observational setups, \textit{HST} WFC3+\textit{Spitzer} IRAC, while the inclusion for a 2nd thermal profile is largely unjustified within a nested Bayesian hypothesis testing framework (e.g., the fits do not improve enough to justify the additional parameter), significant biases in the abundance may exist at large contrasts.  In particular we found that an artificially precise constraint on the methane abundance can be obtained when assuming a hemisphere composed of a single 1D thermal profile.  For a representative \textit{JWST} observational scenario (1-11 microns requiring the NIRISS, NIRCam, and MIRI instruments), we found strong evidence of a 2nd profile in all contrast cases.  While little molecular abundance biases appeared to exist for the lowest contrast (0.2), significant biases exist in the water, carbon monoxide, and ammonia abundances for high contrast (0.8). We also found that the retrieval was able to accurately recover both TP profiles when included in the model. 

\ccolor{Conceptually, we can understand why the 1TP retrieval performs poorly in the case of large contrast by considering just the blackbody spectra of the day and the night sides. Because the night side flux is much lower, the averaged flux we observe is essentially half of the day side flux. This averaged spectrum is then not of a blackbody form. The 1TP approach can be thought of the attempt to fit one blackbody to the averaged spectrum -- it cannot simultaneously fit for both the peak location and the amplitude. An alternative way to fit for the lowered flux, and allowing the fitting of the peak, is to change the emitting area. In our case, that area is fixed, making that not applicable. The 1TP retrieval has to rely on the flexibility provided by tweaking the thermal profile and abundances. With a 2TP approach, we are able to halve one of the blackbodies in the same way the data are generated, and we can better characterize this simple day-night atmosphere.}

As a practical real-world example, we tested the 1 TP vs. 2 TP profile on the first quarter phase, third quarter phase, and day side emission spectra of the hot Jupiter WASP-43b as observed with \textit{HST} WFC3 \citep{stevenson14} and \textit{Spitzer} IRAC (Stevenson et al., in prep). For the dayside, the results are analogous to the low contrast synthetic cases. For the first quarter, we found, much like in our high contrast synthetic model scenarios, that a strong methane bias appears when assuming only a single 1D profile, but that the retrieved water abundance remains robust under the different assumptions.  The artificially strong methane constraint is driven by the requirement to fit the IRAC 3.6$\mu$m point given only a single TP profile to work with, whereas the water abundance constraint is driven primarily by the WFC3 data of which is less impacted by the assumption of one or two TP profiles. The inclusion of a 2nd TP profile in this particular scenario is justified at the moderate to strong 3.3 $\sigma$ level.  

It is prudent for us to note, however, that for WASP-43b vertical mixing could potentially reproduce our single TP scenario retrieved methane abundance ($\sim 10^{-5}$).  The abundance of methane near the base of the single TP profile at typical CH$_4$-CO quench pressures of $\sim$10 bars \citep[e.g.,][1600 K]{Moses11,line2011} is a few $\times 10^{-5}$. So, in a sense, if we assume a single TP profile, we would arrive to the conclusion that the measured methane abundance is indicative of disequilibrium chemistry to a high degree of constraint (i.e., solar composition thermochemical equilibrium would have been ruled out at several sigma in this scenario). Instead, if we assume two TP profiles, the methane upper limit would be consistent with both pure thermochemical equilibrium at solar composition or solar composition with quenched methane.  We are inclined to believe the latter scenario (two profiles) given our synthetic test cases and the fairly strong detection thresh-hold for the 2nd TP profile. The broad methane upper limit permits both chemical situations.  Furthermore,  \citet{Kreidberg14} found only an upper limit on the methane abundance from the day side emission and transmission spectra of WASP-43b. Had disequilibrium methane been as present as it appeared so here, under the single TP assumption, we would have expected a similar, if not higher, degree of constraint on the methane abundance due to the slightly higher signal-to-noise of the feature during occultation.  \ccolor{This WASP-43b example clearly points out a degeneracy in the interpretation of the spectrum, non-equilibrium chemistry or not, which can only be lifted with a robust determination of additional TP profiles that comes from higher S/N spectra over a wider wavelength range.}

For the third quarter, the posterior for methane remains constrained regardless of the retrieval set-up. Instead of a statement on the chemical processes present at this phase, we take this result to highlight future work that should be done to examine the effects of utilizing broadband photometric points and the consistency of retrievals for a full phase curve.

\ccolor{\section{Future Work}} \label{future}
\ccolor{As we continue to push the envelope in exoplanet atmosphere observations, at the cutting edge we will always be trying to make initial inferences about planetary climate and atmospheric abundances from data with limited wavelength ranges and less than ideal signal-to-noise.  Here we have shown that with sparse data, and even with outstanding data over a wide wavelength range, that modeling choices can dramatically impact our view of an atmosphere's retrieved parameters.  In addition to considering and defending choices made for observational strategies and data reduction methods, it would be wise for us to also consider choices made in the construction of our model retrievals.} 

This manuscript serves as an initial investigation of the impact of spatial inhomogeneities on our interpretation of emission spectra.  Much work remains to be explored, including, but not limited to the impacts of:  spatially non-homogeneous molecular abundances driven by disequilibrium processes or instantaneous equilibrium, day side single or multiple ``hot-spots", optically thick non-uniform clouds (like in brown dwarfs), and a more thorough sweep of the observational parameter space (wavelength coverage, signal-to-noise, resolution). 

\ccolor{The exploration of observational set-ups is especially important in the coming years. The current wavelength coverage provided by WFC3+IRAC does not offer the information necessary to differentiate between potentially contrasting profiles.} The \textit{JWST} results show the potential wealth of information at wavelengths not currently probed by space-based observations. For our explored case, with its prescribed abundances and parameters, $2-3\mu$m are essential in highlighting thermal contrast. It will be worthwhile to explore this behavior under different conditions. We also emphasize this characteristic because it demonstrates our ability to determine diagnostic wavelengths indicative of key features in an atmosphere with future observations in mind. Moving forward, we aim to explore how we can minimize the observational coverage needed while maximizing our inference. 

Our investigation, along with the recent exploration of non-uniform terminator cloud cover by \citet{Line15}, serves to demonstrate that there is a strong need to consider the non-homogeneous nature of transiting exoplanets when interpreting their spectra. While there has been a push to develop ever more sophisticated and complicated 1D models, we have shown that even the simplest of 2D assumptions can strongly impact the models, and may even potentially dwarf the impact of the more sophisticate physics being explored in the 1D models. Moving forward, we suggest a balanced approach between complicated 1D models and simple 2D models when interpreting transit (both emission and transmission) spectra. Starting from simple models and working toward more complicated models permits us to better understand the importance of the inclusion of additional model physics \citep[e.g.,][]{showman2011}.

\acknowledgments
We thank Xi Zhang, Caroline Morley, Mark Swain, and Gautam Vasisht for useful and instructive conversations. We also thank Tom Greene for providing us with estimated JWST uncertainties. This material is based upon work supported by the National Science Foundation Graduate Research Fellowship under Grant DGE1339067. The computation for this research
was performed by the UCSC Hyades supercomputer, which is supported by National Science Foundation
(award number AST-1229745) and University of
California, Santa Cruz. M.R.L. acknowledges support provided by NASA through Hubble Fellowship grant 51362 awarded by the Space Telescope Science Institute, which is operated by the Association of Universities for Research in Astronomy, Inc., for NASA, under the contract NAS 5-26555.  J.J.F. acknowledges support of Hubble grant \textit{HST}-GO-13467.03-A, NSF grant A16-0321-001, and NASA grant NNX02AH23G. Based on observations made with the NASA/ESA \textit{Hubble Space Telescope} and the NASA \textit{Spitzer Space Telescope}.

\bibliographystyle{apj}


\end{document}